\documentclass{article}

\usepackage{arxiv}

\usepackage[utf8]{inputenc} % allow utf-8 input
\usepackage[T1]{fontenc}    % use 8-bit T1 fonts
\usepackage{hyperref}       % hyperlinks
\usepackage{url}            % simple URL typesetting
\usepackage{booktabs}       % professional-quality tables
\usepackage{amsfonts}       % blackboard math symbols
\usepackage{nicefrac}       % compact symbols for 1/2, etc.
\usepackage{microtype}      % microtypography
\usepackage{lipsum}
\usepackage{graphicx}
\graphicspath{ {./images/} }

\usepackage{graphicx}
\usepackage[]{todonotes}
\usepackage{url}
\usepackage{siunitx}
\usepackage{tabularx,multirow}    % for \multirow{5}{*}{text}
\usepackage{amsthm,amsmath}
\usepackage{mathtools}
\DeclarePairedDelimiter\abs{\lvert}{\rvert}%
\usepackage[utf8]{inputenc}

\title{Investigating the Performance Gap between Testing on Real and Denoised Aggregates in Non-Intrusive Load Monitoring}

\author{
Christoph Klemenjak \\
  Institute of Networked and Embedded Systems\\
  University of Klagenfurt\\
  Austria \\
  \texttt{klemenjak@ieee.org} \\
   \And
 Stephen Makonin \\
  School of Engineering Science\\
  Simon Fraser University\\
  Canada\\
  \texttt{smakonin@sfu.ca} \\
  \And
 Wilfried Elmenreich \\
  Institute of Networked and Embedded Systems\\
  University of Klagenfurt\\
  Austria \\
  \texttt{wilfried.elmenreich@aau.at} \\
}

\begin{document}
\maketitle
\begin{abstract}
Prudent and meaningful performance evaluation of algorithms is essential for the progression of any research field. In the field of Non-Intrusive Load Monitoring (NILM), performance evaluation can be conducted on real-world aggregate signals, provided by smart energy meters or artificial superpositions of individual load signals (i.e., denoised aggregates). It has long been suspected that testing on these denoised aggregates provides better evaluation results mainly due to the the fact that the signal is less complex. Complexity in real-world aggregate signals increases with the number of unknown/untracked load. Although this is a known performance reporting problem, an investigation in the actual performance gap between real and denoised testing is still pending.
In this paper, we examine the performance gap between testing on real-world and denoised aggregates with the aim of bringing clarity into this matter. Starting with an assessment of noise levels in datasets, we find significant differences in test cases. We give broad insights into our evaluation setup comprising three load disaggregation algorithms, two of them relying on neural network architectures. The results presented in this paper, based on studies covering three scenarios with ascending noise levels, show a strong tendency towards load disaggregation algorithms providing significantly better performance on denoised aggregate signals. A closer look into the outcome of our studies reveals that all appliance types could be subject to this phenomenon. We conclude the paper by discussing aspects that could be causing these considerable gaps between real and denoised testing in NILM. 
\end{abstract}

% keywords can be removed

\section*{Introduction}
Effective energy management in smart grids requires a fair amount of monitoring and controlling of electrical load to achieve optimal energy utilization and, ultimately, reduce energy consumption \cite{gopinath2020energy}. With regard to individual buildings, load monitoring can be implemented in an intrusive or non-intrusive fashion. The latter is often referred to as Non-Intrusive Load Monitoring (NILM) or load disaggregation. NILM, dating back to the seminal work presented in \cite{hart1985prototype}, comprises a set of techniques to identify active electrical appliance signals from the aggregate load signal reported by a smart meter \cite{Salem2020}.

\begin{figure} []
  \begin{minipage}[b]{0.47861\textwidth}
\includegraphics[]{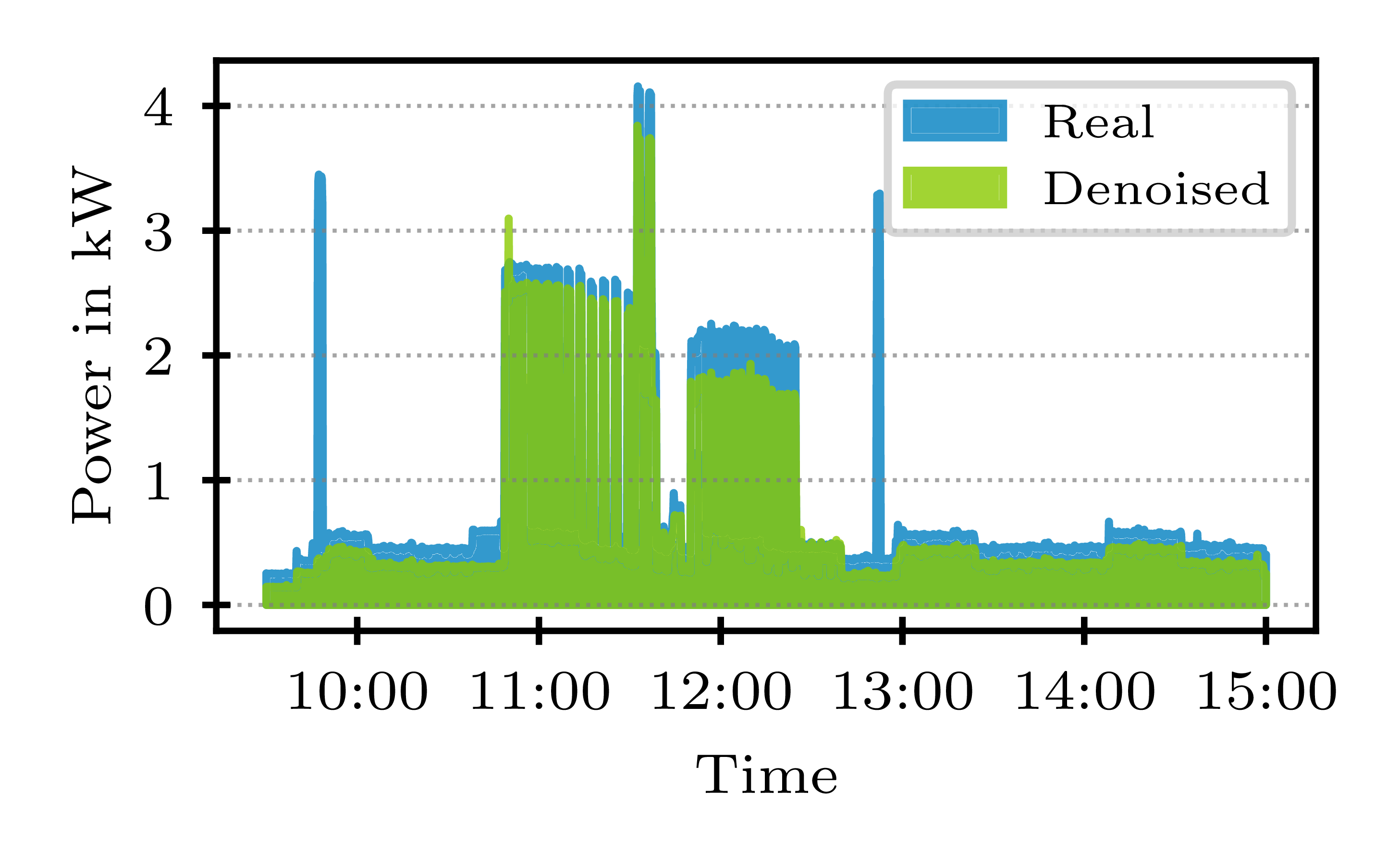}
    \caption{Real and denoised aggregate in the case of UK-DALE house 5}
    \label{fig:aggregates_uk}
  \end{minipage}
  \begin{minipage}[b]{0.47861\textwidth}
\includegraphics[]{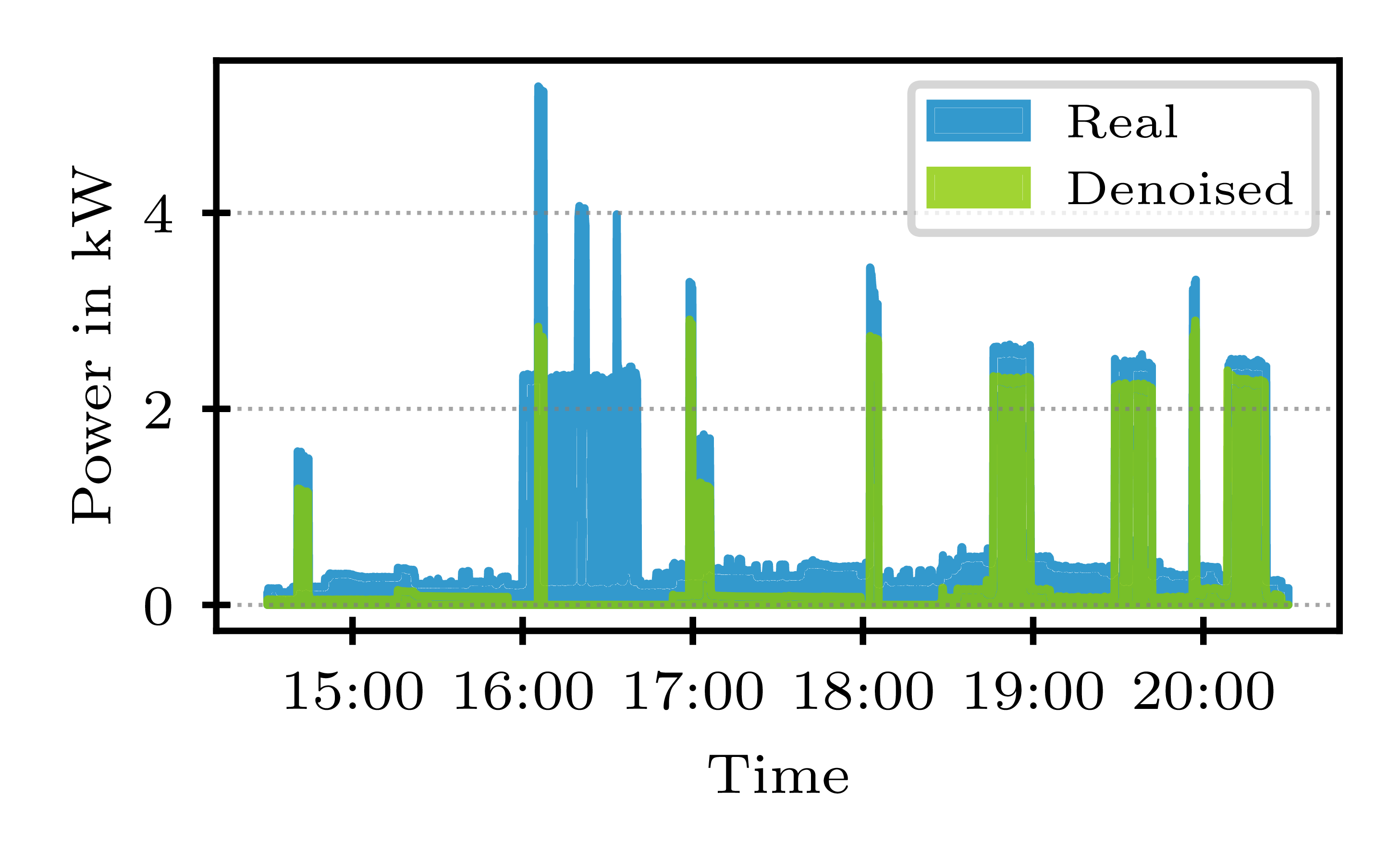}
    \caption{Real and denoised aggregate in the case of REFIT house 2}
    \label{fig:aggregates_refit}
  \end{minipage}
\end{figure}

% lead towards testing and report what has been done....
Performance evaluation of NILM algorithms can be carried out in a noised or denoised manner, where the difference lies in the aggregate signal considered as input. Whereas noised scenarios employ signals (i.e. time series) obtained from smart meters, denoised testing scenarios consider superpositions of individual appliance signals (i.e., denoised aggregates). Figure \ref{fig:aggregates_uk} and Figure \ref{fig:aggregates_refit} illustrate such real and denoised signals for two households found in NILM datasets.

While a large proportion of contributions proposed for NILM is being evaluated following noised testing scenarios, exceptions to this unwritten rule can be observed \cite{wittmann2018nonintrusive}. The problem with this matter lies in the complexity of the test setup, as denoised aggregates are suspected to pose simpler disaggregation problems \cite{makonin2015nonintrusive}. Consequently, our hypothesis claims that the same disaggregation algorithm applied to the denoised signal version of a real-world aggregate signal results in considerably better performance, thus communicating a distorted picture of the capabilities of the presented algorithm. 

% Explain what this paper is about and speak about the outline shortly
This paper presents a study with a focus on the difference of denoised and real-world signal testing scenarios in the context of performance evaluation in NILM. On the basis of test runs considering data of 15 appliances extracted from three datasets with considerably different noise levels, we strive towards bringing clarity on this widely disregarded question. We incorporate one basic as well as two load disaggregation approaches based on neural networks to obtain a broad understanding whether or not noise levels of aggregate power signals impact energy estimation performance. Finally, we discuss how the disaggregation performance is affected by signal noise levels with regard to different appliance types.

\section*{Related Work} \label{sec:related}

Despite the possibly far-reaching implications of this aspect for NILM, relatively little is understood about the actual performance gap between real and denoised testing. In \cite{makonin2015nonintrusive}, the hypothesis of denoised testing resulting in better performance was expressed first. Further, the authors introduce a measure to assess the noise level of aggregate signals. This measure has found application in a limited number of studies, in which the noise level was reported alongside the performance of load disaggregation algorithms on real-world aggregates \cite{makonin2015exploiting}, \cite{zhao2018improving}. However, no comparison to the denoised testing case has been conducted. In \cite{klemenjak2020towards}, the noise levels of several NILM datasets were determined. The authors report basic parameters of several NILM datasets and find that noise levels in real aggregate signals vary significantly among observed datasets. 

%\todo[]{TODO: Bonfigli did very similar things while evaluating his DAE in bonfigli2018denoising. We have to acknowledge that carefully while not reducing the importance of our contribution... }

Few attempts have been made to evaluate NILM algorithms on both, real and denoised aggregates, such as presented for the AFAMAP approach in \cite{bonfigli2017non}. In subsequent work \cite{bonfigli2018denoising}, an improved version of denoising autoencoders for NILM has been proposed by means of comparison studies to the state of the art at that time. Although the authors have not investigated the performance gap between real and denoised, a tendency can be derived for this particular case in both contributions, confirming the motivation for the studies presented in this paper.

\section*{Assessing Signal Noise Levels} \label{sec_ch3_datanoise}

% Explain what NILM is and introduce NAR
NILM has been approached in a variety of ways that can be categorized into event detection and energy estimation approaches \cite{pereira2018performance}. In this investigation, we put an emphasis on the energy estimation viewpoint, as it can be seen as the precursor of the event detection stage in the disaggregation process. We define NILM as the problem of generating estimates $[\hat{x}_t^{(1)}, \dots ,\hat{x}_t^{(M)}]$ of the actual power consumption $[x_t^{(1)}, \dots ,x_t^{(M)}]$ of $M$ electrical appliances at time $t$ given only the aggregated power consumption $y_t$, where the aggregate power signal $y_t$ consists of
\begin{equation}
	y_t =  \sum_{i=1}^{M}{x_t^{(i)}}+ \eta_t
\end{equation}
that is $M$ appliance-level signals $x_t^{(i)}$ and a residual term $\eta_t$. The residual term comprises (measurement) noise as well as the sum of unmetered electrical load \cite{klemenjak2020towards}. To quantify the share of unmetered load in an aggregate signal, the noise-aggregate ratio NAR, defined as:
\begin{equation}
	\text{NAR} = \frac{\sum_{t=1}^{T}{\eta_t}}{\sum_{t=1}^{T}{y_t}} = \frac{\sum_{t=1}^{T}{|y_t-\sum_{i=1}^{M}{x_t^{(i)}}| }}{\sum_{t=1}^{T}{y_t}}
\end{equation}
was introduced in \cite{makonin2015nonintrusive}. This ratio can be computed for any type of power signal, provided that readings of the aggregate and individual appliances are available. A NAR of 0.15 reports that 15\% of the observed power signal can be attributed to the residual term. Hence, the ratio indicates to what degree information on the aggregate's components is available.

To get an impression of NAR levels to be expected in real-world settings, we compute this ratio for households embedded in the energy datasets AMPds2~\cite{makonin2016ampds}, COMBED~\cite{batra2014comparison}, ECO~\cite{beckel2014eco}, iAWE~\cite{batra2013s}, REFIT~\cite{murray2017electrical}, and UK-DALE~\cite{kelly2015uk}. These datasets were selected because of their compatibility to NILMTK, a toolkit that enables reproducible NILM experiments \cite{batra2019towards, batra2014nilmtk}. We excluded from consideration the dataset BLUED \cite{anderson_blued_2012} due to the lack of sub-metered power data, Tracebase \cite{reinhardt2012accuracy} and GREEND \cite{monacchi2014greend} due to the lack of household aggregate power data. 
We summarize the derived values in Table \ref{tab_ch3_noise_in_datasets} in conjunction with further stats on the measurement campaign such as duration or number of submeters. 

\begin{table}[]
\caption{Noise levels in NILM datasets}
\vspace*{3mm}
\small
\centering
\renewcommand{\arraystretch}{1.1}
\begin{tabular}{ccccccc}
\hline
\textbf{Dataset}              & \textbf{House}                & \textbf{Duration}                       & \textbf{Meters}               & \multicolumn{2}{c}{\textbf{Power Types}}   & \textbf{NAR}   \\ 

%\multicolumn{1}{l}{} & \multicolumn{1}{l}{} & \multicolumn{1}{c}{} & \multicolumn{1}{c}{} & \multicolumn{1}{c}{\textbf{}} & \multicolumn{1}{c}{\textbf{}} & \multicolumn{1}{c}{\textbf{P}} & \multicolumn{1}{c}{\textbf{S}}  \\ 
\multicolumn{1}{l}{} & \multicolumn{1}{l}{} & \multicolumn{1}{c}{{[}days{]}} & \multicolumn{1}{c}{} & \multicolumn{1}{c}{\textbf{}} & \multicolumn{1}{c}{\textbf{}} & \multicolumn{1}{c}{{[}\%{]}}  \\
 \hline
AMPds2       &  1     &  730          &   20     &     P, Q, S           &   P, Q, S          &     17.8              \\ %\rowcolor[HTML]{EFEFEF} 

COMBED      &  1     &  28          &   13     &     P           &   P          &     34.4            \\

ECO       &  1     &  236          &   7     &     P, Q           &   P          &     67.0                 \\

ECO       &  2     &  245          &   12     &     P, Q           &   P          &     5.9                \\ %\rowcolor[HTML]{EFEFEF} 

ECO       &  3     &  57          &   7     &     P, Q           &   P          &     97.0                  \\

ECO       &  4     &  211          &   8     &     P, Q           &   P          &     70.5                 \\ %\rowcolor[HTML]{EFEFEF} 

ECO       &  5     &  219          &   8     &     P, Q           &   P          &     84.7                  \\
ECO       &  6     &  124          &   7     &     P, Q           &   P          &     66.0                 \\ %\rowcolor[HTML]{EFEFEF} 
iAWE       &  1     &  60          &   10     &     P, Q, S           &   P, Q, S          &     50.0            \\   

%RAE       &  1     &  -          &   -     &    P           &   P          &     -       & -            \\ %\rowcolor[HTML]{EFEFEF} 

%RAE       &  2     &  -          &   -     &     P           &   P          &     -       & -            \\  

REFIT       &  1     &  639          &   9     &    P           &   P          &     64.5                  \\ %\rowcolor[HTML]{EFEFEF} 
REFIT       &  2     &  617          &   9     &     P           &   P          &     65.1             \\ 
 
REFIT & 3   &  614          &   9     &  P          &   P       & 55.5         \\ 

REFIT & 4    &  634          &   9     &  P          &   P       & 52.5            \\

REFIT & 5     &  648          &   9     &  P          &   P      & 52.3         \\  

UK-DALE       &  1     &  658          &   52     &     P, S           &   P, S          &     33.3                  \\ %\rowcolor[HTML]{EFEFEF} 

UK-DALE       &  2     &  110          &   18     &     P, S           &   P          &     41.2                  \\

UK-DALE       &  3     &  35          &   4     &     S           &   P          &     -             \\%\rowcolor[HTML]{EFEFEF} 

UK-DALE       &  4     &  114          &   5     &    S            &    P         &     -               \\ 

UK-DALE       &  5     &  107          &   24     &    P, S            &    P         &     27.5                 \\ \hline
\end{tabular}
\label{tab_ch3_noise_in_datasets}
\end{table}

Generally speaking, measurement campaigns strive to record the energy consumption and other parameters of interest in households or industrial facilities over a certain time period. Though sharing this common aim, considerable differences can be observed in the way past campaigns have been conducted. As Table \ref{tab_ch3_noise_in_datasets} shows, durations range from a couple of days to several years of data, which impacts the amount of appliance activations and events found in the final dataset. Further, we identify considerable variations with regard to AC power types as well as the number of submeters installed during campaigns. It should be pointed out that there seems to be a lack of consistency in the sense that not only measurement setups differ between two datasets but also within some of the campaigns considered by our comparison (e.g., UK-DALE).

As concerns the noise aggregate ratio (NAR), we observe considerable variations across datasets and households. Interestingly, the NAR ranges between a few percent, as it is the case for household 2 in the ECO dataset, and excessive 84.7\% in household 5 of same dataset. Further, there are indications that the number of submeters used in the course of dataset collection can but do not necessarily have an impact on the noise level of the household's aggregate signal since it is decisive what kind of appliances are left out during a measurement campaign (low-power appliances vs. big consumers). As concerns house 1 to house 5 in REFIT, we consistently observe moderate to high noise levels, which may be the result of the low number of submeters incorporated in the measurement campaign. 
On the other hand, it should be noted that the measurement campaign conducted to obtain REFIT shows remarkable consistency in the sense that the exact same number of submeters has been applied to every single household in the study and, more importantly, the same AC power type has been considered at aggregate and appliance level at every site.
In contrast to that, Table \ref{tab_ch3_noise_in_datasets} reveals that in the case of house 3 and 4 in UK-DALE, apparent power was recorded on aggregate level, whereas active power was considered on appliance level only. As our definition of NAR demands for the same AC power type on aggregate and submeter level, no such ratio could be computed in those cases. The same applies to all sites of the REDD \cite{kolter2011redd} dataset, according to the NILMTK dataset converter\footnote{\scriptsize \url{https://github.com/nilmtk/nilmtk/tree/master/nilmtk/dataset_converters/redd/metadata}}. For this reason, REDD has not been considered in this study.

\section*{Evaluation Setup} \label{sec_ch3_study}

%\subsection*{Evaluation Setup} \label{sec_ch3_setup}

To gain a comprehensive understanding of the impact of noise on the disaggregation performance of algorithms, we selected three households with ascending levels of residual noise: household 2 of the ECO dataset \cite{beckel2014eco} with a NAR of 5.9\%, household 5 of the UK-DALE dataset \cite{kelly2015uk} with a NAR 27.5\%, and household 2 of the REFIT dataset \cite{murray2017electrical} with a NAR of 65.1 \%. 
This way, we incorporate one instance each for disaggregation problems with low, moderate, and high noise levels. For every household considered, we selected five electrical appliances and spent best efforts to consider a wide range of appliance types. We extracted 244 days for ECO, 82 days for UK-DALE and 275 days for REFIT while applying a sampling interval of \SI{10}{\second}. The amount of data used per household was governed by availability in the case of ECO and UK-DALE, as can be learned from Table \ref{tab_ch3_noise_in_datasets}. As concerns REFIT, the considered time window was March 1st to December 1st of 2014. We split datasets into training set (70\%), validation set (15\%), and test set (15\%). This splitting was applied to all three households.
We considered the smart meter signal as present in datasets and obtained the denoised version of the aggregate by superposition of the individual appliance signals following:
\begin{equation}
	y_t =  \sum_{i=1}^{M}{x_t^{(i)}}
\end{equation}
For experimental evaluations, we utilize the latest version of NILMTK. The toolkit integrates several basic benchmark algorithms as well as load disaggregation algorithms based on Deep Neural Networks (DNN). In the course of experiments, we consider the traditional CO approach and two approaches based on DNNs:
\begin{itemize}
\item The \emph{Combinatorial Optimization (CO)} algorithm, introduced in \cite{hart1985prototype}, has been used repeatedly in literature to serve as baseline \cite{batra2019towards}. The CO algorithm estimates the power demand of appliances and their operational mode. Similar to the Knapsack problem~\cite{rodriguez-silva_unilm}, estimation is performed by finding the combination of concurrently active appliances that minimizes the difference between aggregate signal and the sum of power demands.
    
\item \emph{Recurrent Neural Networks} are a subclass of neural networks that have been developed to process time series and related sequential data \cite{dipietro2020deep}. First proposed for NILM in~\cite{kelly15neuralnilm}, we employ the implementation presented in \cite{krystalakos18windowgru}, which incorporates Long Short-Term Memory (LSTM) cells. Provided a sequence of aggregate readings as input, the RNN estimates the power consumption of the electrical appliance it was trained to detect for each newly observed input sample.
    
\item The \emph{Sequence-to-point (S2P)} technique, relying on convolutional neural networks, follows a sliding window approach in which the network predicts the midpoint element of an output time window based on an input sequence consisting of aggregate power readings~\cite{zhang2018sequence}. The basic idea behind this method is to implement a non-linear regression between input window and midpoint element, which has been applied successfully for speech and image processing \cite{oord2016wavenet}. In a recent benchmarking study of NILM approaches, S2P was observed to be amongst the most advanced disaggregation techniques at that time \cite{reinhardt2020eenergy}.
\end{itemize}

While the CO approach does not need to be parametrized, we set the number of training epochs to 25 during training of neural networks. Further, we employ an input sequence length of 49 for LSTM inspired by \cite{krystalakos18windowgru} and 99 for S2P as suggested in \cite{batra2019towards}.

In this study, we utilize two error metrics to assess the performance of load disaggregation algorithms. The first is a well-known, common metric used in signal processing, the Mean Absolute Error (MAE), defined as:
\begin{equation}\label{eq:mae}
    \text{MAE}^{(i)} = \frac{1}{T} \cdot \sum_{t=1}^{T}{\abs*{ \hat{x}_t^{(i)}-x_t^{(i)}}}  
\end{equation}
where $x_t$ is the the actual power consumption, $\hat{x}_t$ the estimated power consumption, and $T$ represents the number of samples. The best possible value is zero and, as we estimate the power consumption of appliances, it is measured in Watts.
As second metric, we incorporate a metric defined by NILM scholars in \cite{kolter2012approximate}, the Normalized Disaggregation Error (NDE), defined as:
\begin{equation}
\text{NDE}^{(i)} =  \sqrt{\frac{\sum_{t=1}^{T}{(\hat{x}_t^{(i)}-x_t^{(i)}})^2}{\sum_{t=1}^{T}{(x_t^{(i)}})^2}}
\end{equation}
In contrast to the MAE, the NDE represents a dimensionless metric and, more importantly, the NDE belongs to the class of normalized metrics. This allows for fair comparisons of disaggregation performance between appliance types \cite{klemenjak2020towards}.

\section*{Results} 
\begin{table}[]
\caption{Mean Absolute Error (MAE) in Watts for real and denoised testing}
\vspace*{3mm}
\small
\renewcommand{\arraystretch}{1.1}
        \centering
\begin{tabular}{ccrcccccccc}
\hline
   &    &      & \multicolumn{2}{c}{\textbf{CO}}        &   & \multicolumn{2}{c}{\textbf{LSTM}}          &                                                              & \multicolumn{2}{c}{\textbf{S2P}}                                     \\ \cline{4-5} \cline{7-8} \cline{10-11} 
     &   &            \textbf{Appliance}                    & \textbf{Real}    & \textbf{Den}                          & & \textbf{Real}      & \textbf{Den}          &              & \textbf{Real}      & \textbf{Den}                         \\ \hline
 
&&audio system       & 37.7 & 32.3 && 6.4 & 5.6 && 6.8 & 5.9 \\

&&dishwasher       & 43.1 & 40.2 && 7.5 & 3.9 && 5.8 & 3.6 \\

&&fridge       & 41.8 & 49.8 && 9.5 & 11.7 && 7.5 & 8.5 \\

%&&HTPC     &  & 56.6      & 27.4      &  & 16.7       & 6.4       &  & 15.6      & 6.5     &  & 0.86     & 0.64     &  & 0.34   & 0.16      &  & 0.32       & 0.16      \\

&&kettle    & 17.3 & 42.7 && 4.2 & 2.5 && 3.2 & 1.3 \\

\parbox[t]{2mm}{\multirow{-5}{*}{\rotatebox[origin=c]{90}{ECO (2)}}}&  \parbox[t]{2mm}{\multirow{-5}{*}{\rotatebox[origin=c]{90}{\footnotesize NAR = 5.9 \%}}}& lamp       & 62.2 & 47.1 && 28.9 & 16.4 && 28.4 & 16.5 \\ \hline
 
& &dishwasher      & 87.6 & 95.1 && 12.2 & 3.6 && 8.3 & 4.1 \\
&&electric oven      & 50.7 & 39.2 && 20.7 & 9.0 && 17.4 & 9.1 \\
&&electric stove      & 131.3 & 40.1 && 7.3 & 6.1 && 6.4 & 4.4 \\
&&fridge     & 161.4 & 141.4 && 25.7 & 21.1 && 20.6 & 16.4 \\

\parbox[t]{2mm}{\multirow{-5}{*}{\rotatebox[origin=c]{90}{UK-DALE (5)}}}&  \parbox[t]{2mm}{\multirow{-5}{*}{\rotatebox[origin=c]{90}{\footnotesize NAR = 27.5 \%}}}& washing machine  & 74.8 & 51.4 && 28.6 & 14.8 && 17.3 & 13.7 \\

\hline
 
&&dishwasher     & 96.0 & 41.3 && 31.4 & 9.0 && 25.3 & 8.5 \\
&&fridge       & 57.8 & 22.8 && 23.0 & 10.4 && 23.9 & 12.4 \\
&&kettle       & 79.1 & 9.2 && 9.8 & 3.3 && 9.7 & 3.2 \\
&&microwave       & 70.8 & 46.6 && 2.7 & 1.5 && 2.8 & 1.1 \\
\parbox[t]{2mm}{\multirow{-5}{*}{\rotatebox[origin=c]{90}{REFIT (2)}}}&  \parbox[t]{2mm}{\multirow{-5}{*}{\rotatebox[origin=c]{90}{\footnotesize NAR = 65.1\%}}}& washing machine      & 101.5 & 41.0 && 21.6 & 12.6 && 24.1 & 11.3 \\
\hline
\end{tabular}
%\vspace*{4mm}
      % \caption{Mean absolute error for real and denoised testing}
       \label{tab_sec_3_MAE}

\end{table}
\begin{table}[]
      \caption{Normalised Disaggregation Error (NDE) for real and denoised testing}
      \vspace*{3mm}
\small
\renewcommand{\arraystretch}{1.1}
        \centering
\begin{tabular}{ccrcccccccc}
\hline
   &    &      & \multicolumn{2}{c}{\textbf{CO}}        &   & \multicolumn{2}{c}{\textbf{LSTM}}          &                                                              & \multicolumn{2}{c}{\textbf{S2P}}                                     \\ \cline{4-5} \cline{7-8} \cline{10-11} 
     &   &            \textbf{Appliance}                    & \textbf{Real}    & \textbf{Den}                          & & \textbf{Real}      & \textbf{Den}          &              & \textbf{Real}      & \textbf{Den}                         \\ \hline
 
&&audio system       & 1.83 & 1.74 && 0.48 & 0.42 && 0.47 & 0.44 \\

&&dishwasher       & 0.96 & 0.78 && 0.44 & 0.26 && 0.34 & 0.22 \\

&&fridge       & 1.8 & 2.07 && 0.45 & 0.5 && 0.38 & 0.36 \\

%&&HTPC     &  & 56.6      & 27.4      &  & 16.7       & 6.4       &  & 15.6      & 6.5     &  & 0.86     & 0.64     &  & 0.34   & 0.16      &  & 0.32       & 0.16      \\

&&kettle    & 1.3 & 1.66 && 0.51 & 0.34 && 0.48 & 0.22 \\

\parbox[t]{2mm}{\multirow{-5}{*}{\rotatebox[origin=c]{90}{ECO (2)}}}&  \parbox[t]{2mm}{\multirow{-5}{*}{\rotatebox[origin=c]{90}{\footnotesize NAR = 5.9 \%}}}& lamp       & 1.27 & 1.18 && 0.74 & 0.5 && 0.74 & 0.52 \\ \hline
 
& &dishwasher      & 1.44 & 1.55 && 0.76 & 0.39 && 0.61 & 0.34 \\
&&electric oven      & 1.51 & 0.98 && 0.66 & 0.38 && 0.53 & 0.33 \\
&&electric stove      & 2.9 & 1.82 && 0.66 & 0.58 && 0.6 & 0.42 \\
&&fridge     & 3.16 & 3.24 && 0.64 & 0.6 && 0.55 & 0.46 \\
\parbox[t]{2mm}{\multirow{-5}{*}{\rotatebox[origin=c]{90}{UK-DALE (5)}}}&  \parbox[t]{2mm}{\multirow{-5}{*}{\rotatebox[origin=c]{90}{\footnotesize NAR = 27.5 \%}}}& washing machine  & 1.47 & 1.16 && 0.63 & 0.35 && 0.42 & 0.32 \\

\hline
 
&&dishwasher     & 1.06 & 0.74 && 0.56 & 0.19 && 0.48 & 0.18 \\
&&fridge       & 1.4 & 1.81 && 0.7 & 0.48 && 0.68 & 0.48 \\
&&kettle       & 1.33 & 0.43 && 0.48 & 0.2 && 0.46 & 0.2 \\
&&microwave       & 4.54 & 3.84 && 0.85 & 0.45 && 0.84 & 0.36 \\
\parbox[t]{2mm}{\multirow{-5}{*}{\rotatebox[origin=c]{90}{REFIT (2)}}}&  \parbox[t]{2mm}{\multirow{-5}{*}{\rotatebox[origin=c]{90}{\footnotesize NAR = 65.1\%}}}& washing machine      & 2.04 & 1.42 && 0.82 & 0.5 && 0.75 & 0.45 \\
\hline
\end{tabular}
%\vspace*{3mm}
 
       \label{tab_sec_3_NDE}
\end{table}

We summarize the outcome of our investigations in Table \ref{tab_sec_3_MAE} for the MAE and Table \ref{tab_sec_3_NDE} with regard to the NDE. For several appliances per household, we compare the disaggregation performance of CO, LSTM, and S2P when applied to the real-world aggregate signal, denoted as \emph{Real}, and the denoised aggregate signal \emph{Den}, respectively.

% describe what Table 2 reports; strong tendency towards denoised being easier; though some exceptions were observed
In virtually all cases, we observe a strong tendency towards disaggregation algorithms providing better performance on denoised aggregate signals. In the context of error metrics such as MAE and NDE this means that the error observed on the real aggregate is larger than the error on the denoised aggregate. This holds true for almost all households and appliances considered, though some exceptions were identified: 
we spot a few cases in Table \ref{tab_sec_3_MAE}, namely the fridge and kettle in ECO as well as the dishwasher in UK-DALE showing the opposite trend for the CO algorithm. Same applies to all fridges with regard to the NDE metric, as Table \ref{tab_sec_3_NDE} reports. It should be pointed out that in those cases, the performance of CO on the real-world and denoised aggregate signal shows a considerable gap when compared to LSTM and S2P. Therefore and because of CO being a trivial benchmarking algorithm, we claim that these cases can be neglected.

As concerns LSTM and S2P, we identify a single contradictory observation, namely in the case of the fridge in ECO's household 2. In this particular case, we observed that testing on the real-world aggregate signal results in marginally better performance. One explanation for this could be the extremely low NAR in this scenario, 5.9\%, and the fridge belonging to the category of appliances with a recurrent pattern \cite{reinhardt2020eenergy}.

% talk about the trends depicted by the bar graphs...
Having identified a clear tendency towards CO, LSTM, and S2P providing significantly better performance in the denoised signal case i.e. lower MAE and NDE, we draw our attention to the open question whether or not there exists a link between noise level and the magnitude of the performance gap between \emph{Real} and \emph{Den}. To investigate further in this, we define the performance gap to be the distance between the error on the real aggregate signal and the error observed signal when testing on the denoised aggregate signal:
\begin{equation}
    \Delta\text{MAE} = \text{MAE}_{\text{real}} - \text{MAE}_{\text{denoised}}
\end{equation}
\begin{equation}
    \Delta\text{NDE} = \text{NDE}_{\text{real}} - \text{NDE}_{\text{denoised}}
\end{equation}

\begin{figure} []
  \begin{minipage}[b]{0.47861\textwidth}
\includegraphics[]{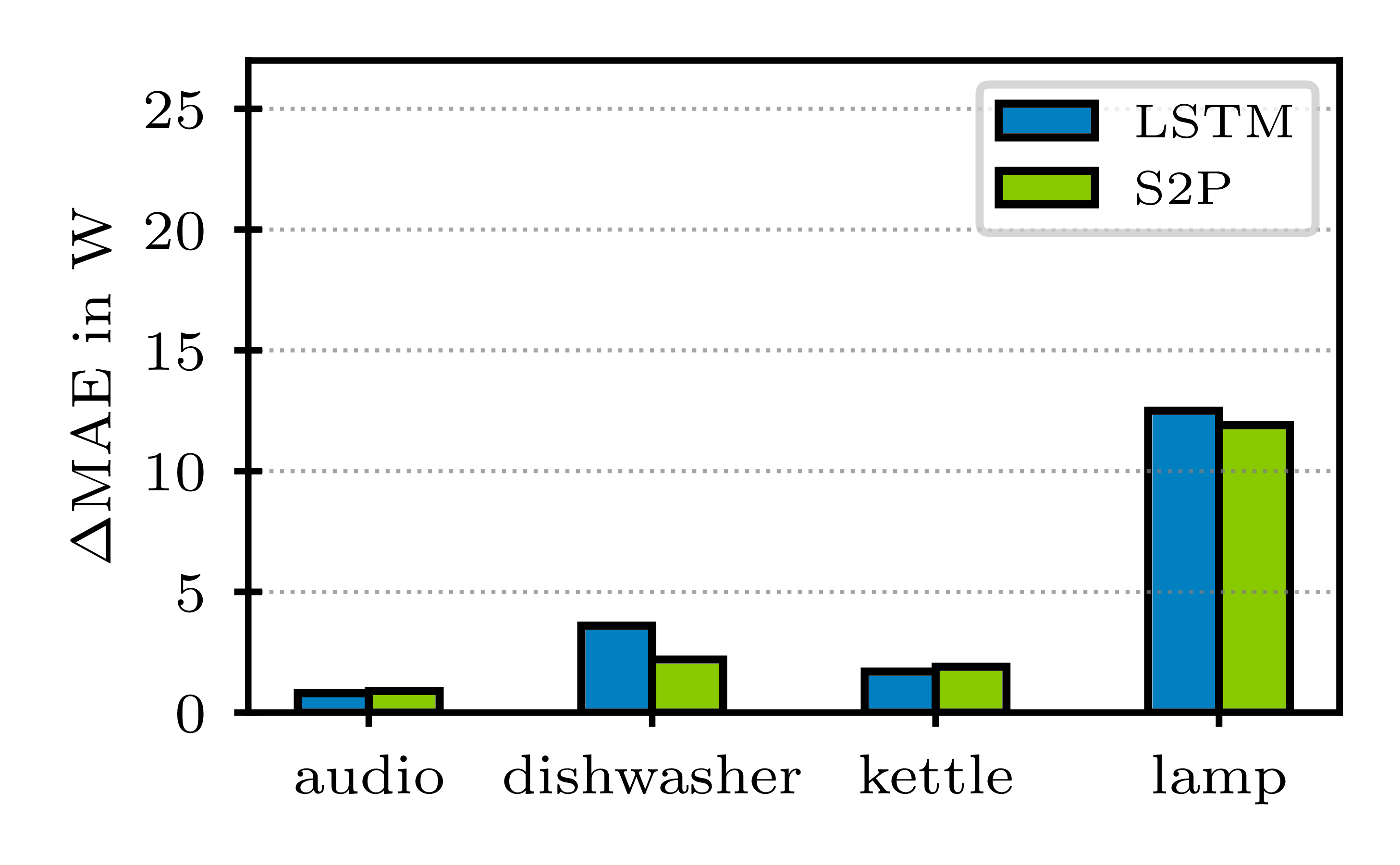}
    \caption{Performance gap with regard to MAE for ECO house 2}
    \label{fig:gap_mae_eco}
  \end{minipage}
\end{figure}

\begin{figure} []
  \begin{minipage}[b]{0.47861\textwidth}
\includegraphics[]{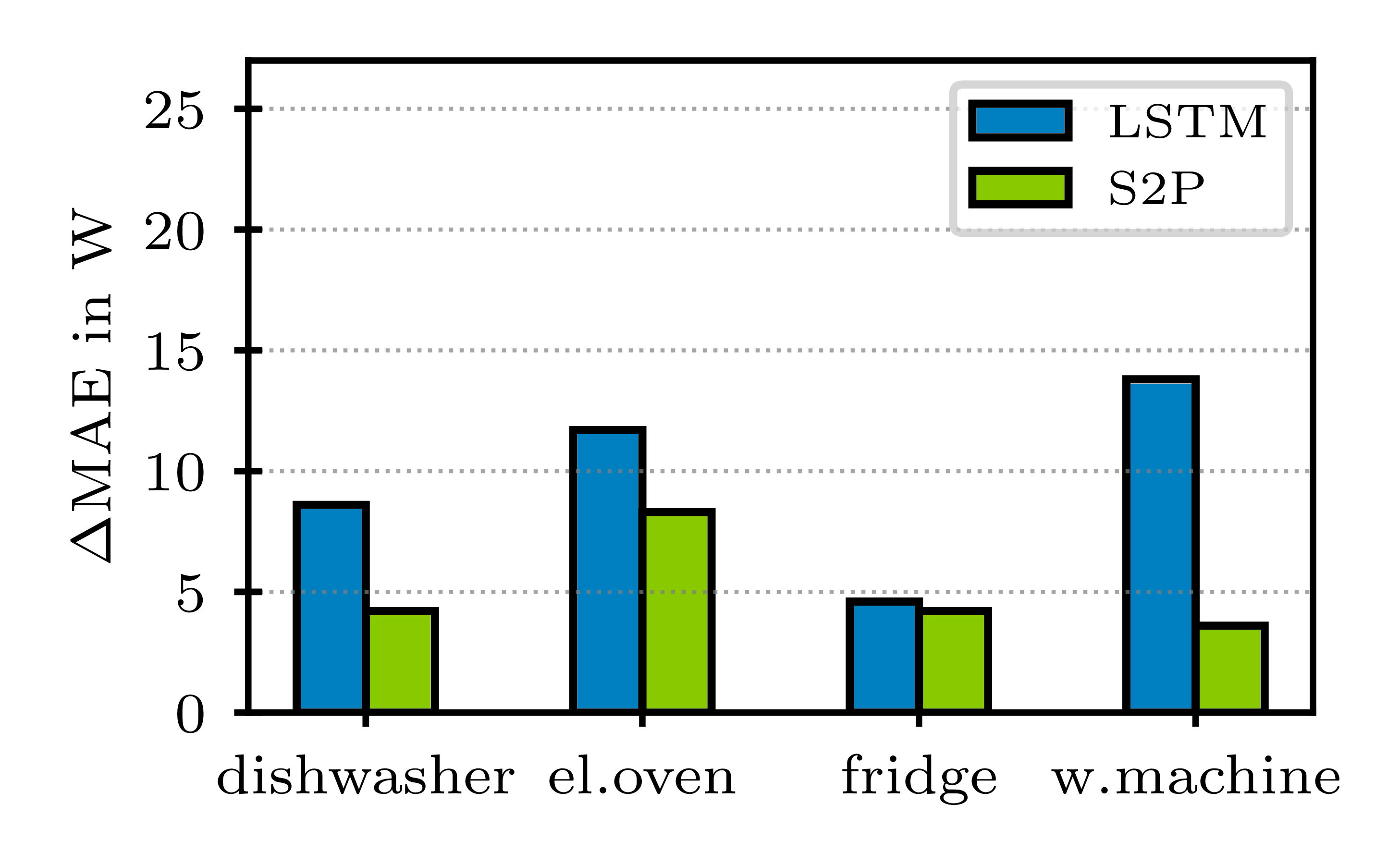}
    \caption{Performance gap with regard to MAE for UK-DALE house 5}
    \label{fig:gap_mae_uk}
  \end{minipage}
  \begin{minipage}[b]{0.47861\textwidth}
\includegraphics[]{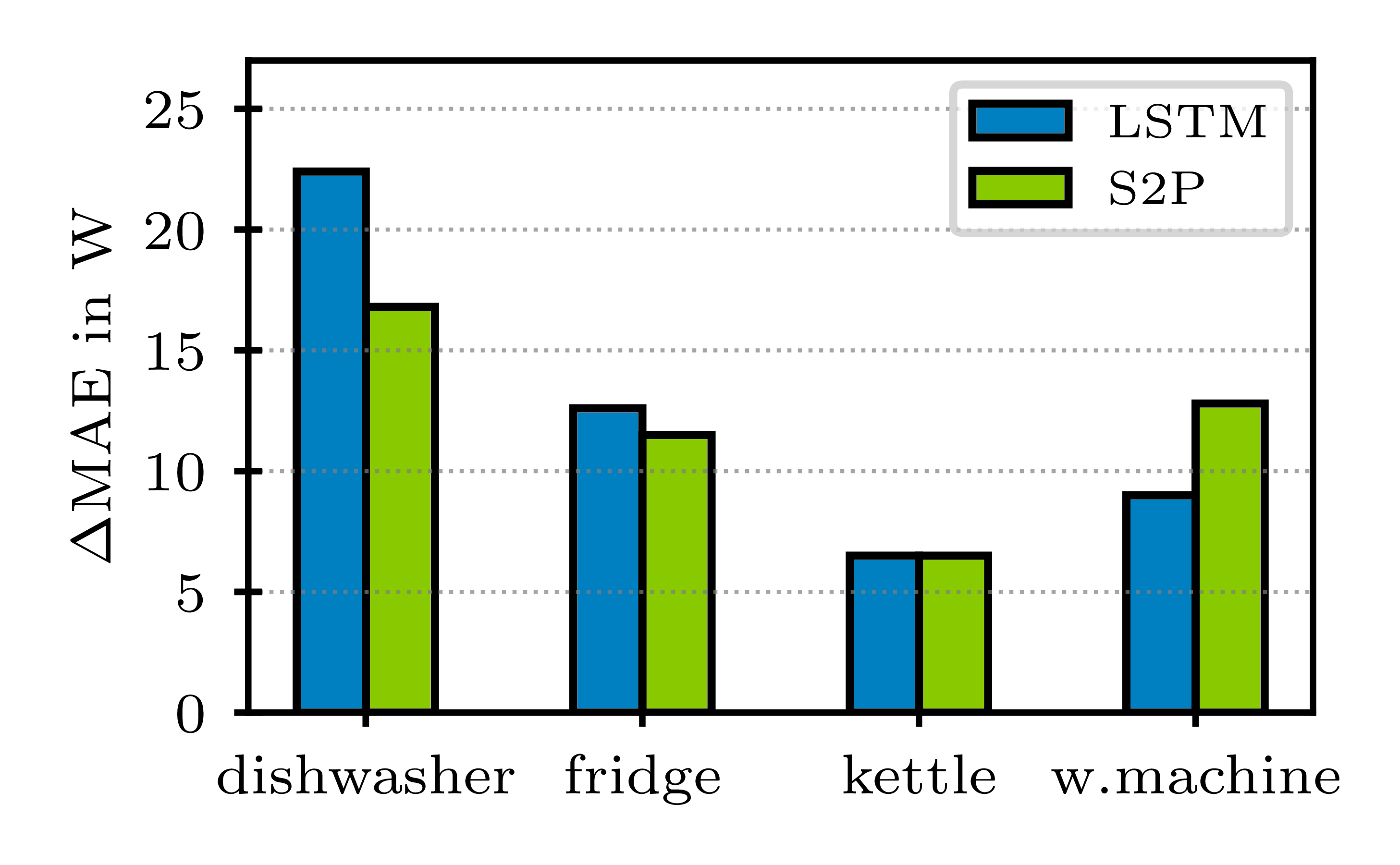}
    \caption{Performance gap with regard to MAE for REFIT house 2}
    \label{fig:gap_mae_refit}
  \end{minipage}
\end{figure}

We derive $\Delta\text{MAE}$ for the cases presented in Table \ref{tab_sec_3_MAE} and illustrate an excerpt of found gaps in Figure \ref{fig:gap_mae_eco} for ECO, Figure \ref{fig:gap_mae_uk} for UK-DALE, and Figure \ref{fig:gap_mae_refit} for REFIT, where the focus of this discussion lies on the two approaches based on neural networks.

We observe clear gaps for both NILM approaches based on neural nets, LSTM and S2P. The illustrations show that neither approach seems to be resilient to noise. This is particularly interesting as approaches relying on LSTM cells as well as sequence-to-sequence learning have received increased interest lately \cite{reinhardt2020eenergy, kaselimi2019bayesian, kaselimi2020context, mauch2015new, wang2019nonintrusive}.
Further, we identify higher performance gaps in test cases on REFIT's house 2 compared to house 5 of UK-DALE in this study. This is particularly apparent when comparing the performance gap for the dishwasher across households, where we measure a $\Delta\text{MAE}$ many times higher in case of REFIT. Also, we observe performance gaps twice as high for the fridge on REFIT compared to UK-DALE. The only exception to this trend represents the case of LSTM for washing machines, where the performance gap of the LSTM network is smaller on REFIT than on UK-DALE.

Nevertheless, it should be stressed that comparisons based on not-normalized metrics can, but not have to be, misleading in some cases since two appliances of the same kind (i.e., two dishwashers) may differ significantly in terms of power consumption. Furthermore, metrics are designed to measure specific aspects of algorithms and hence, considering several metrics during performance evaluation results in a broader understanding of the capabilities of algorithms.

\begin{figure} []
  \begin{minipage}[b]{0.47861\textwidth}
\includegraphics[]{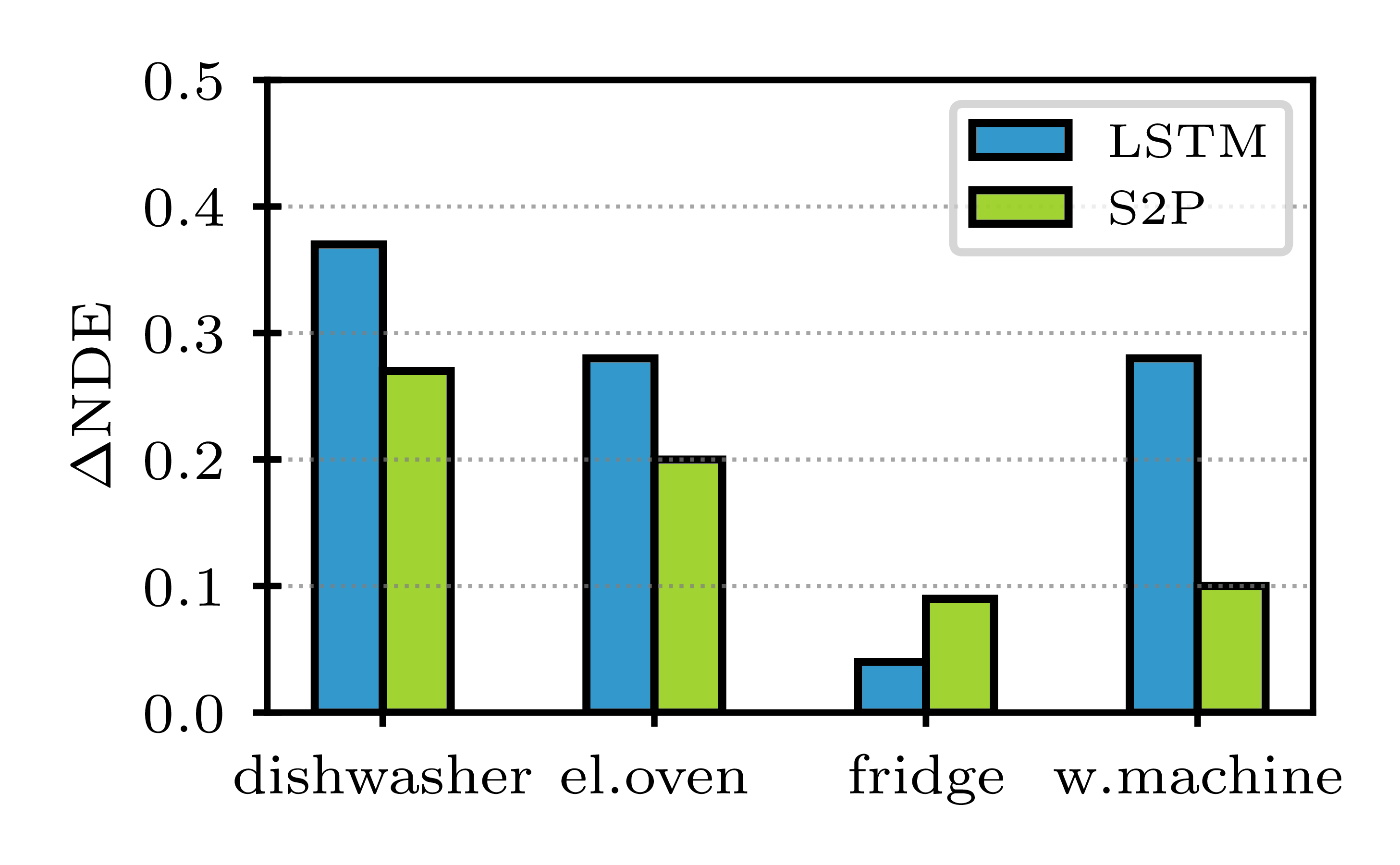}
    \caption{Performance gap with regard to NDE for UK-DALE house 5}
    \label{fig:gap_nde_uk}
  \end{minipage}
  \begin{minipage}[b]{0.47861\textwidth}
\includegraphics[]{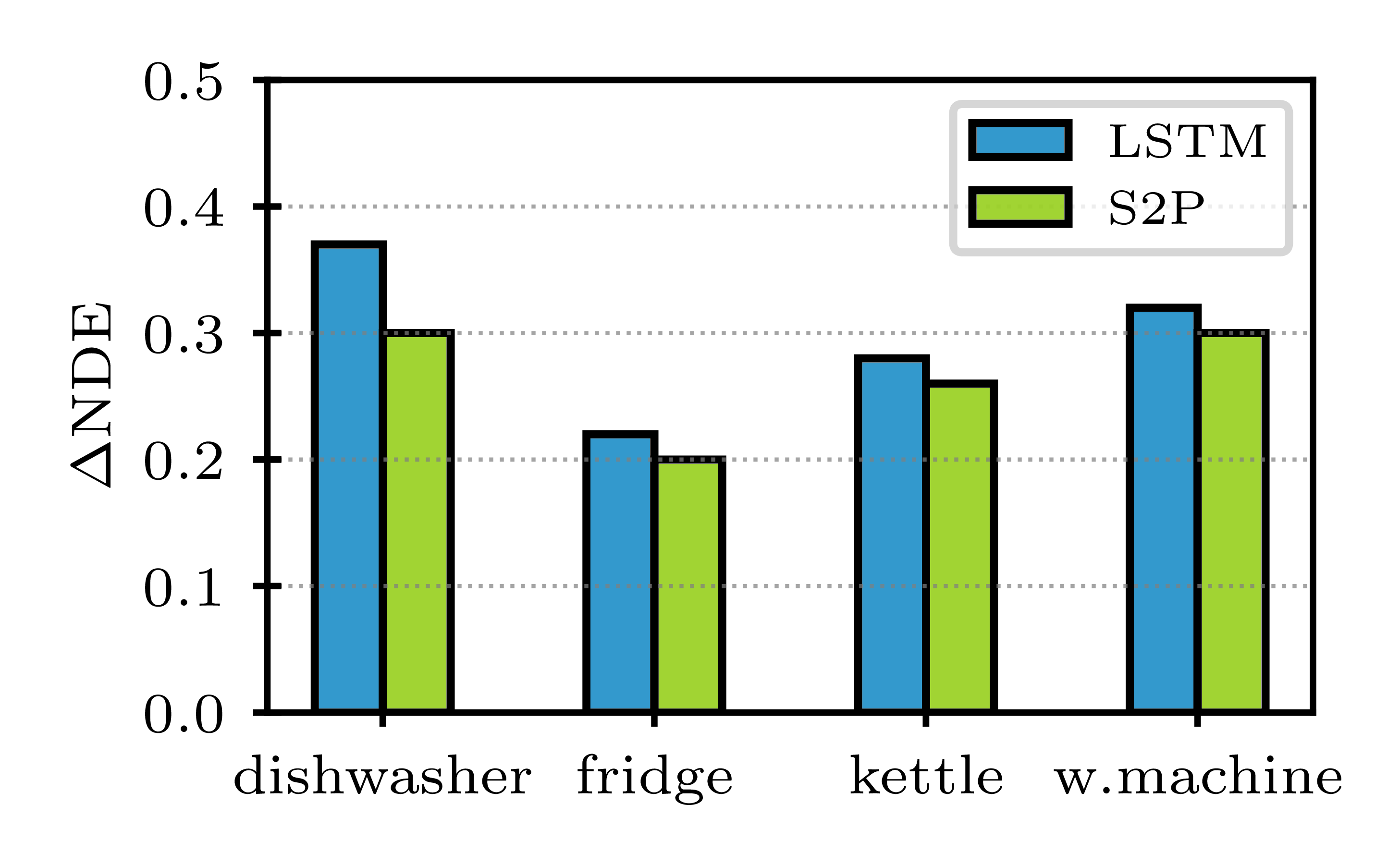}
    \caption{Performance gap with regard to NDE for REFIT house 2}
    \label{fig:gap_nde_refit}
  \end{minipage}
\end{figure}

For these reasons, we also derived performance gaps with regard to NDE, $\Delta\text{NDE}$, for the test cases presented in Table \ref{tab_sec_3_NDE} and illustrate derived gaps in Figure \ref{fig:gap_nde_uk} for UK-DALE and Figure \ref{fig:gap_nde_refit} for REFIT.

In the case of fridges, we observe substantially lower performance gaps on UK-DALE for both networks. We suspect that is a result of the comparably high amount of noise in REFIT 2, disaggregating the real-world aggregate signal represents a bigger challenge than in the case of the denoised counterpart, especially when estimating the power consumption of low-power household appliances such as fridges.

\begin{figure} []
  \begin{minipage}[b]{0.47861\textwidth}
\includegraphics[]{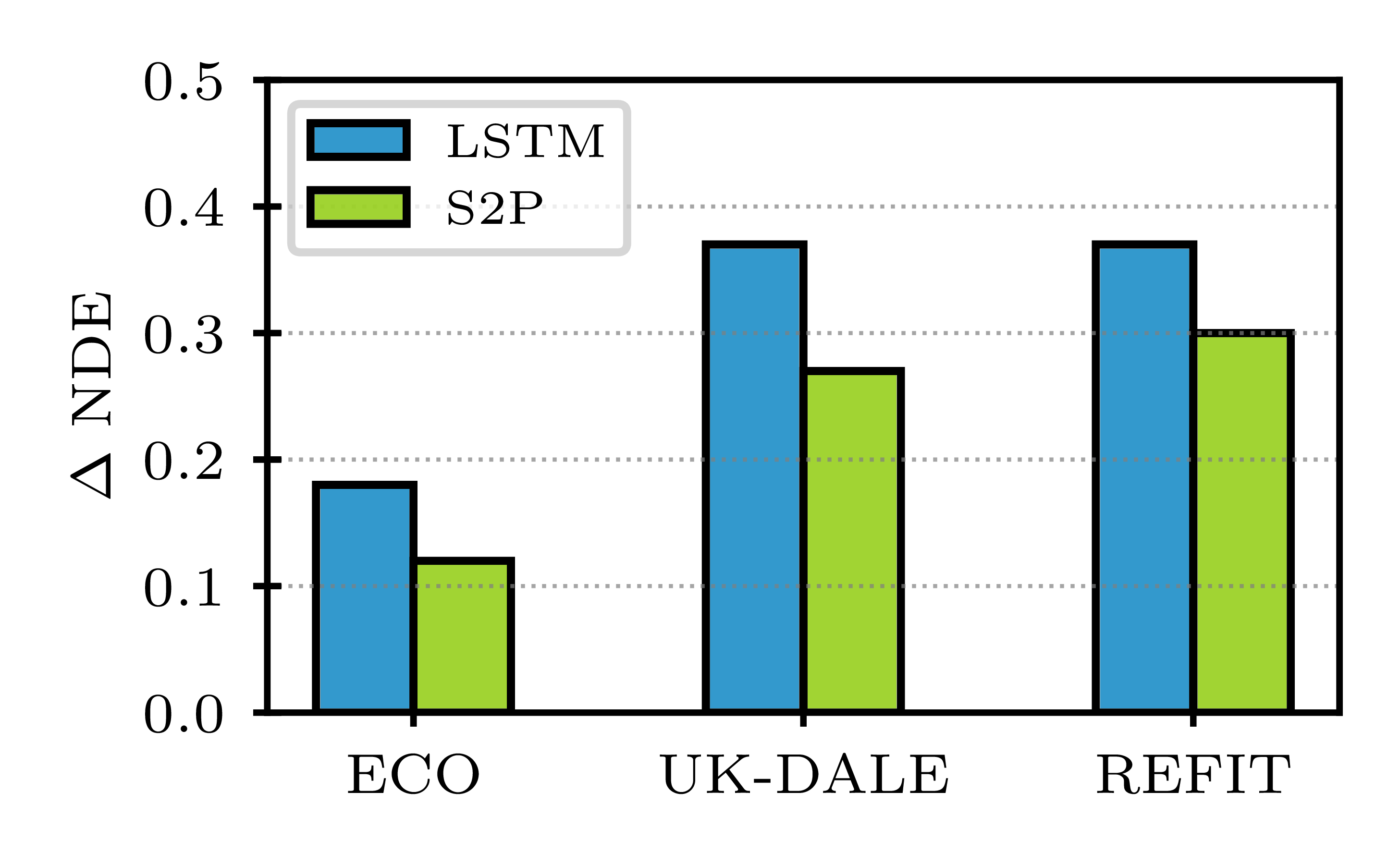}
    \caption{Performance gap with regard to NDE for dishwashers}
    \label{fig:gap_dishwashers}
  \end{minipage}
  \begin{minipage}[b]{0.47861\textwidth}
\includegraphics[]{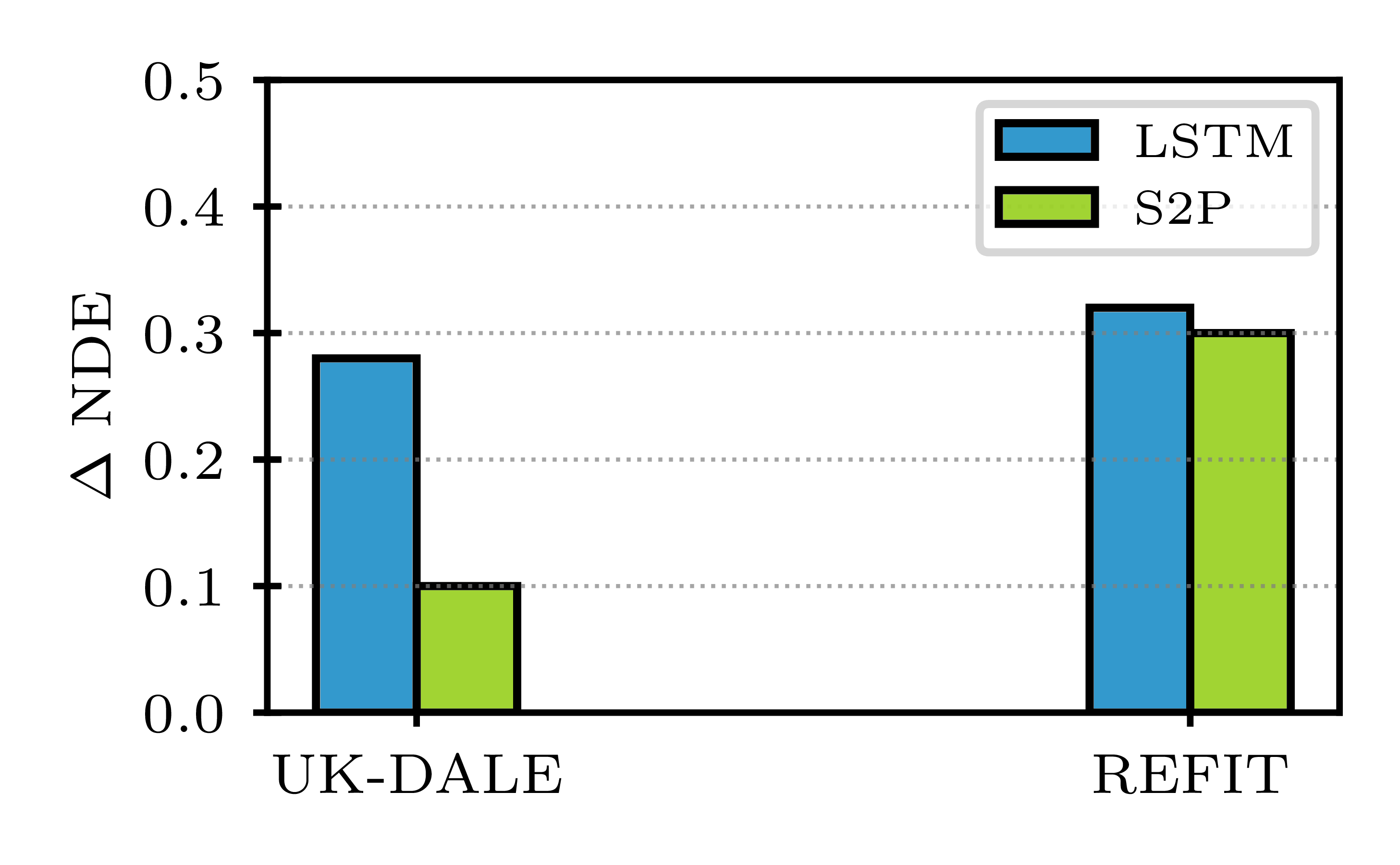}
    \caption{Performance gap with regard to NDE for washing machines}
    \label{fig:gap_washing}
  \end{minipage}
\end{figure}

Interestingly, not only we observe considerable performance gaps when estimating the power consumption of low-power appliances but also for appliances with moderate or high power consumption such as dishwashers and washing machines, as can be learned from Figure \ref{fig:gap_dishwashers} and Figure \ref{fig:gap_washing}. In both cases, UK-DALE and REFIT, we measure the highest $\Delta\text{NDE}$ in the case of the dishwasher. A comparison of performance gaps for dishwashers in Figure \ref{fig:gap_dishwashers} reveals that while we measure similar performance gaps in UK-DALE and REFIT, the performance gap in the case of ECO is significantly smaller. We hypothesize this is the result of the marginal noise level measured in house 2 of ECO. More importantly, we observe that also in cases of marginal noise levels, an apparent difference in terms of disaggregation error can be observed between real and denoised testing in this example.

A recent benchmarking study involving eight disaggregation algorithms found that S2P outperformed competing neural network architectures and concluded that S2P ranks amongst the most promising NILM approaches \cite{reinhardt2020eenergy}. 
As concerns performance of NILM algorithms interpreted as disaggregation error between estimated power consumption and true power consumption of appliances, we find that S2P outperforms LSTM in 11 of 15 cases for the MAE metric and in 14 of 15 cases when the NDE metric is considered.
Furthermore, in the vast majority of test runs, the S2P approach shows lower performance gaps than the network composed of LSTM cells in the sense of $\Delta\text{MAE}$ and $\Delta\text{NDE}$.

\section*{Discussion} \label{sec_ch3_discussion}
\begin{figure}[ht]
\centering
\includegraphics[]{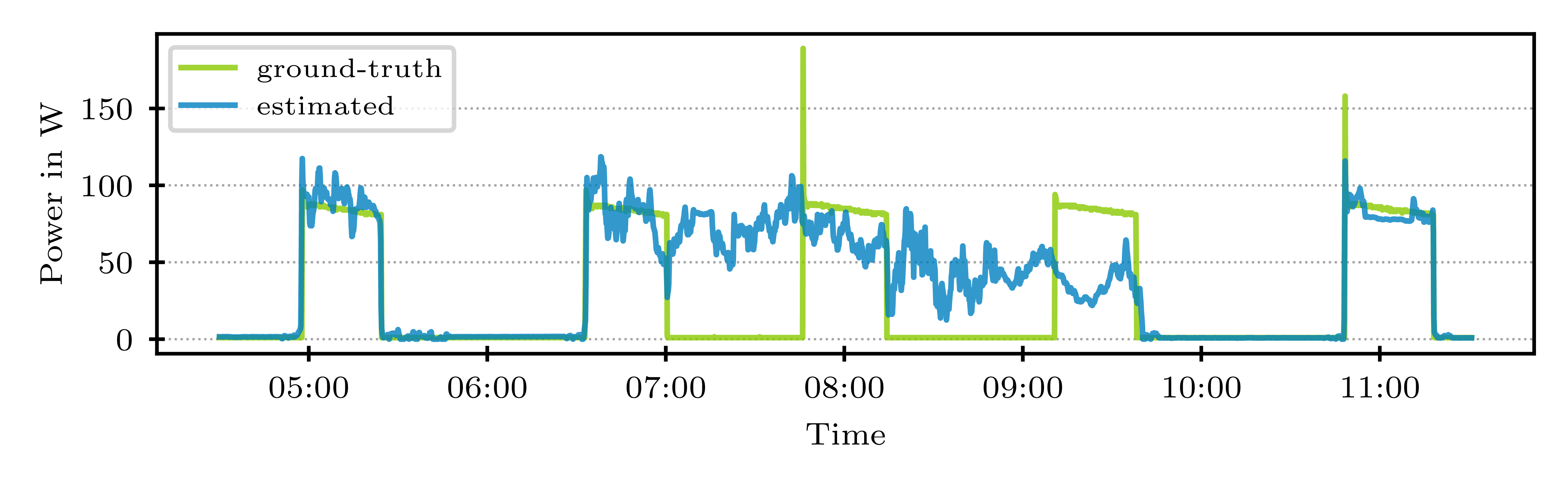}
\caption{An excerpt of estimates provided by S2P for the fridge in REFIT house 2 when applied to the real aggregate.}
\label{fig:refit-signal-real}
\end{figure}

\begin{figure}[ht]
\centering
\includegraphics[]{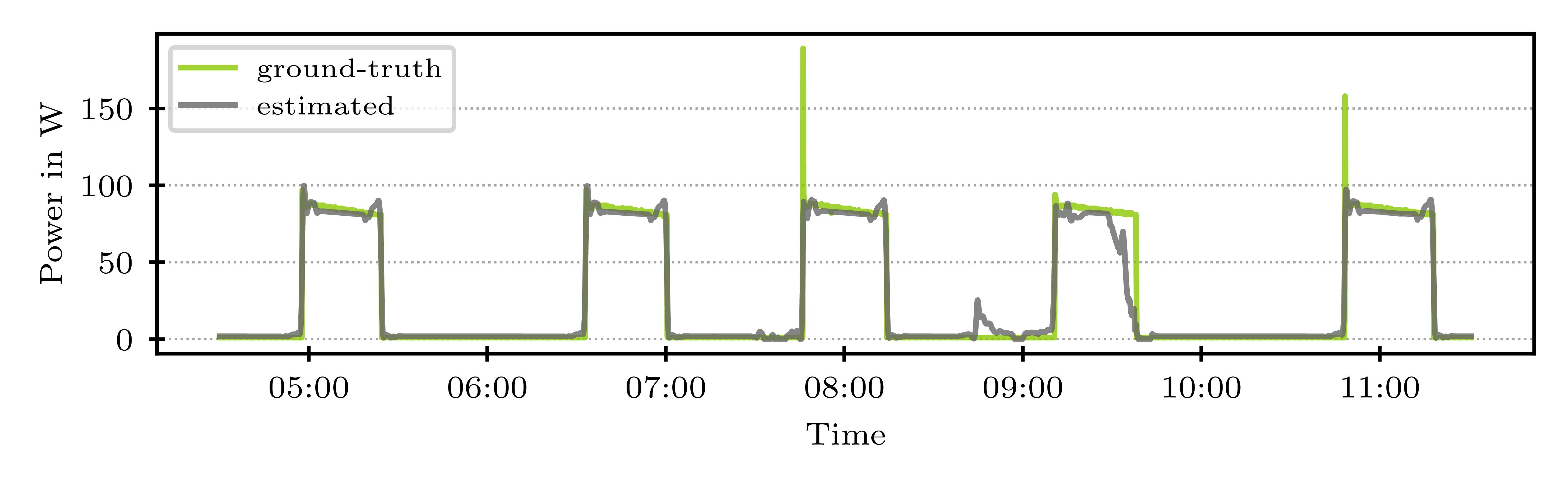}
\caption{An excerpt of estimates provided by S2P for the fridge in REFIT house 2 when applied to the denoised aggregate.}
\label{fig:refit-signal-den}
\end{figure}

Insights obtained from testing on three households with considerably different NAR levels reveal that in the majority of test runs, testing on the denoised aggregate signal leads to substantially lower estimation errors and therefore, higher estimation accuracy. A few cases showing the contrary trend were observed but can be reasonably explained. As this apparent performance gap can be attributed to a variety of aspects, we suspect two of them having a decisive impact on this matter:

First, denoised aggregates are obtained by superposition of individual appliance signals. As such, they \emph{contain fewer appliance activations and consumption patterns} than aggregates obtained from smart meters, respectively. Particularly when estimating the power consumption of low-power appliances, such activations have the potential to hinder load disaggregation algorithms from providing accurate power consumption estimates. Such cases were repeatedly observed during our studies on REFIT, where a NAR of 65.1 \% was measured. As depicted in Figure \ref{fig:refit-signal-real} and Figure \ref{fig:refit-signal-den}, we detected several cases where concurrent operation of appliances with moderate or high power consumption (i.e. dishwasher, electric stove, or washing machine) resulted in significant deviations when estimating the power consumption of the fridge. Not only we observed such cases for the basic benchmarking algorithm CO but also for the advanced NILM approaches RNN and S2P, which leads to the presumption that though having seen remarkable advances in the the state of the art, at least a part of those algorithms may still be prone to noise levels in aggregate signals.

Second, we observe a substantially \emph{higher number of false positive estimates} in predictions based on real-world aggregate signals than in estimates generated from denoised aggregate signals. False positives in this context mean that the NILM algorithms predicted the appliance to consume energy at times this was not the case. Such false positives impact the outcome of performance evaluations two-fold, as they increase the disaggregation error and decrease the estimation accuracy of NILM algorithms, respectively. 
We observed repeatedly that in the real-world case, the number of false-positive estimates is considerably higher than in the denoised case. We presume that those false positives are the result of algorithms confusing appliances with similar power consumption levels.

Based on the insights gained in this study, we can, however, not confirm a clear link between noise level, measured in NAR, and the magnitude of the performance gap between testing on real and denoised aggregates. We suspect this is due to the fact that every load disaggregation problem bears individual challenges to load disaggregation algorithms, making a comparison between moderate and high noise levels cumbersome. Though such a positive correlation between noise level and the magnitude of the performance gap could not be confirmed by our evaluation, we demonstrated that it has to be expected that testing on denoised aggregates results in lower disaggregation errors in the majority of test runs. Yet, we would like to stress the need for further investigation into the complexity of load disaggregation problems.

\section*{Conclusions}

Motivated by the use of both, real and denoised aggregates in the evaluation of NILM algorithms in related work, we have investigated the performance gap observed between artificial sums of individual signals and signals obtained from real power meters. First, we utilized a noise measure, the noise-aggregate ratio NAR, to determine the noise level of real-world aggregate signals found in energy datasets. We find that noise levels vary substantially between households. We give insights on the experimental setup employed in our studies, comprising one basic and two more advanced NILM algorithms applied to data from three households with ascending noise levels. Our results show that in virtually all evaluation runs, a significant performance gap between the real and the denoised signal testing case can be identified, provided a sufficiently high noise-aggregate ratio. Though some exceptions were observed, those cases can be well explained. 
Hence, we claim that testing on denoised aggregate signals can lead to a distorted image of the actual capabilities of load disaggregation algorithms in some cases, and ideally, its application should be well-considered when developing algorithms for real-world settings.

\bibliographystyle{unsrt}



\begin{thebibliography}{10}

\bibitem{gopinath2020energy}
R~Gopinath, Mukesh Kumar, C~Prakash~Chandra Joshua, and Kota Srinivas.
\newblock Energy management using non-intrusive load monitoring
  techniques-state-of-the-art and future research directions.
\newblock {\em Sustainable Cities and Society}, page 102411, 2020.

\bibitem{hart1985prototype}
George~W. Hart.
\newblock {Prototype Nonintrusive Appliance Load Monitor}.
\newblock Technical report, MIT Energy Laboratory and Electric Power Research
  Institute, 1985.

\bibitem{Salem2020}
Hajer Salem, Moamar Sayed-Mouchaweh, and Moncef Tagina.
\newblock {\em A Review on Non-intrusive Load Monitoring Approaches Based on
  Machine Learning}, pages 109--131.
\newblock Springer International Publishing, Cham, 2020.

\bibitem{wittmann2018nonintrusive}
Fernando~Marcos Wittmann, Juan~Camilo L{\'o}pez, and Marcos~J Rider.
\newblock Nonintrusive load monitoring algorithm using mixed-integer linear
  programming.
\newblock {\em IEEE Transactions on Consumer Electronics}, 64(2):180--187,
  2018.

\bibitem{makonin2015nonintrusive}
Stephen Makonin and Fred Popowich.
\newblock Nonintrusive load monitoring (nilm) performance evaluation.
\newblock {\em Energy Efficiency}, 8(4):809--814, 2015.

\bibitem{makonin2015exploiting}
Stephen Makonin, Fred Popowich, Ivan~V Baji{\'c}, Bob Gill, and Lyn Bartram.
\newblock Exploiting hmm sparsity to perform online real-time nonintrusive load
  monitoring.
\newblock {\em IEEE Transactions on Smart Grid}, 7(6):2575--2585, 2015.

\bibitem{zhao2018improving}
B.~{Zhao}, K.~{He}, L.~{Stankovic}, and V.~{Stankovic}.
\newblock Improving event-based non-intrusive load monitoring using graph
  signal processing.
\newblock {\em IEEE Access}, 6:53944--53959, 2018.

\bibitem{klemenjak2020towards}
Christoph Klemenjak, Stephen Makonin, and Wilfried Elmenreich.
\newblock Towards comparability in non-intrusive load monitoring: on data and
  performance evaluation.
\newblock {\em 2020 IEEE Power \& Energy Society Innovative Smart Grid
  Technologies Conference (ISGT)}, 2020.

\bibitem{bonfigli2017non}
Roberto Bonfigli, Emanuele Principi, Marco Fagiani, Marco Severini, Stefano
  Squartini, and Francesco Piazza.
\newblock Non-intrusive load monitoring by using active and reactive power in
  additive factorial hidden markov models.
\newblock {\em Applied Energy}, 208:1590--1607, 2017.

\bibitem{bonfigli2018denoising}
Roberto Bonfigli, Andrea Felicetti, Emanuele Principi, Marco Fagiani, Stefano
  Squartini, and Francesco Piazza.
\newblock Denoising autoencoders for non-intrusive load monitoring:
  improvements and comparative evaluation.
\newblock {\em Energy and Buildings}, 158:1461--1474, 2018.

\bibitem{pereira2018performance}
Lucas Pereira and Nuno Nunes.
\newblock Performance evaluation in non-intrusive load monitoring: Datasets,
  metrics, and tools - a review.
\newblock {\em Wiley Interdisciplinary Reviews: Data Mining and Knowledge
  Discovery}, 8(6), 2018.

\bibitem{makonin2016ampds}
Stephen Makonin, Bradley Ellert, Ivan~V. Bajic, and Fred Popowich.
\newblock {Electricity, water, and natural gas consumption of a residential
  house in Canada from 2012 to 2014}.
\newblock {\em Scientific Data}, 3(160037):1--12, 2016.

\bibitem{batra2014comparison}
Nipun Batra, Oliver Parson, Mario Berges, Amarjeet Singh, and Alex Rogers.
\newblock A comparison of non-intrusive load monitoring methods for commercial
  and residential buildings.
\newblock {\em arXiv:1408.6595}, 2014.

\bibitem{beckel2014eco}
Christian Beckel, Wilhelm Kleiminger, Romano Cicchetti, Thorsten Staake, and
  Silvia Santini.
\newblock The eco data set and the performance of non-intrusive load monitoring
  algorithms.
\newblock {\em Proceedings of the 1st ACM Conference on Embedded Systems for
  Energy-Efficient Buildings}, pages 80--89, 2014.

\bibitem{batra2013s}
Nipun Batra, Manoj Gulati, Amarjeet Singh, and Mani~B Srivastava.
\newblock It's different: Insights into home energy consumption in india.
\newblock In {\em Proceedings of the 5th ACM Workshop on Embedded Systems For
  Energy-Efficient Buildings}, pages 1--8, 2013.

\bibitem{murray2017electrical}
David Murray, Lina Stankovic, and Vladimir Stankovic.
\newblock An electrical load measurements dataset of united kingdom households
  from a two-year longitudinal study.
\newblock {\em Scientific data}, 4(1):1--12, 2017.

\bibitem{kelly2015uk}
Jack Kelly and William Knottenbelt.
\newblock The uk-dale dataset, domestic appliance-level electricity demand and
  whole-house demand from five uk homes.
\newblock {\em Scientific data}, 2(1):1--14, 2015.

\bibitem{batra2019towards}
Nipun Batra, Rithwik Kukunuri, Ayush Pandey, Raktim Malakar, Rajat Kumar,
  Odysseas Krystalakos, Mingjun Zhong, Paulo Meira, and Oliver Parson.
\newblock Towards reproducible state-of-the-art energy disaggregation.
\newblock In {\em Proceedings of the 6th ACM International Conference on
  Systems for Energy-Efficient Buildings, Cities, and Transportation}, pages
  193--202, 2019.

\bibitem{batra2014nilmtk}
Nipun Batra, Jack Kelly, Oliver Parson, Haimonti Dutta, William Knottenbelt,
  Alex Rogers, Amarjeet Singh, and Mani Srivastava.
\newblock Nilmtk: an open source toolkit for non-intrusive load monitoring.
\newblock {\em Proceedings of the 5th international conference on Future energy
  systems}, pages 265--276, 2014.

\bibitem{anderson_blued_2012}
Kyle Anderson, Adrian Ocneanu, Diego Benitez, Derrick Carlson, Anthony Rowe,
  and Mario Berges.
\newblock {BLUED:} a fully labeled public dataset for {Event-Based}
  {Non-Intrusive} load monitoring research.
\newblock In {\em Proceedings of the 2nd {KDD} Workshop on Data Mining
  Applications in Sustainability {(SustKDD)}}, Beijing, China, August 2012.

\bibitem{reinhardt2012accuracy}
Andreas Reinhardt, Paul Baumann, Daniel Burgstahler, Matthias Hollick, Hristo
  Chonov, Marc Werner, and Ralf Steinmetz.
\newblock On the accuracy of appliance identification based on distributed load
  metering data.
\newblock In {\em 2012 Sustainable Internet and ICT for Sustainability
  (SustainIT)}, pages 1--9. IEEE, 2012.

\bibitem{monacchi2014greend}
Andrea Monacchi, Dominik Egarter, Wilfried Elmenreich, Salvatore D'Alessandro,
  and Andrea~M Tonello.
\newblock Greend: An energy consumption dataset of households in italy and
  austria.
\newblock {\em 2014 IEEE International Conference on Smart Grid Communications
  (SmartGridComm)}, pages 511--516, 2014.

\bibitem{kolter2011redd}
J.~Zico Kolter and Matthew~J. Johnson.
\newblock Redd: A public data set for energy disaggregation research.
\newblock In {\em Workshop on Data Mining Applications in Sustainability
  (SIGKDD), San Diego, CA}, volume~25, pages 59--62, 2011.

\bibitem{rodriguez-silva_unilm}
A.~{Rodriguez-Silva} and S.~{Makonin}.
\newblock {Universal Non-Intrusive Load Monitoring (UNILM) Using Filter
  Pipelines, Probabilistic Knapsack, and Labelled Partition Maps}.
\newblock In {\em 2019 IEEE PES Asia-Pacific Power and Energy Engineering
  Conference (APPEEC)}, pages 1--6, 2019.

\bibitem{dipietro2020deep}
R~Di~Pietro and GD~Hager.
\newblock Handbook of medical image computing and computer assisted
  intervention.
\newblock {\em Chapter}, 21:503--519, 2019.

\bibitem{kelly15neuralnilm}
Jack Kelly and William Knottenbelt.
\newblock Neural {NILM}: Deep neural networks applied to energy disaggregation.
\newblock In {\em Proceedings of the 2nd ACM International Conference on
  Embedded Systems for Energy-Efficient Built Environments (BuildSys)}, 2015.

\bibitem{krystalakos18windowgru}
Odysseas Krystalakos, Christoforos Nalmpantis, and Dimitris Vrakas.
\newblock Sliding window approach for online energy disaggregation using
  artificial neural networks.
\newblock In {\em Proceedings of the 10th Hellenic Conference on Artificial
  Intelligence (SETN)}, 2018.

\bibitem{zhang2018sequence}
Chaoyun Zhang, Mingjun Zhong, Zongzuo Wang, Nigel Goddard, and Charles Sutton.
\newblock Sequence-to-point learning with neural networks for non-intrusive
  load monitoring.
\newblock In {\em Proceedings of the 32nd AAAI Conference on Artificial
  Intelligence (AAAI)}, 2018.

\bibitem{oord2016wavenet}
Aaron van~den Oord, Sander Dieleman, Heiga Zen, Karen Simonyan, Oriol Vinyals,
  Alex Graves, Nal Kalchbrenner, Andrew Senior, and Koray Kavukcuoglu.
\newblock Wavenet: A generative model for raw audio.
\newblock {\em arXiv preprint arXiv:1609.03499}, 2016.

\bibitem{reinhardt2020eenergy}
Andreas Reinhardt and Christoph Klemenjak.
\newblock How does load disaggregation performance depend on data
  characteristics? insights from a benchmarking study.
\newblock In {\em Proceedings of the Eleventh ACM International Conference on
  Future Energy Systems}, e-Energy `20, pages 167--177, New York, NY, USA,
  2020. Association for Computing Machinery.

\bibitem{kolter2012approximate}
J~Zico Kolter and Tommi Jaakkola.
\newblock Approximate inference in additive factorial hmms with application to
  energy disaggregation.
\newblock In {\em Artificial intelligence and statistics}, pages 1472--1482,
  2012.

\bibitem{kaselimi2019bayesian}
Maria Kaselimi, Nikolaos Doulamis, Anastasios Doulamis, Athanasios Voulodimos,
  and Eftychios Protopapadakis.
\newblock Bayesian-optimized bidirectional lstm regression model for
  non-intrusive load monitoring.
\newblock In {\em ICASSP 2019-2019 IEEE International Conference on Acoustics,
  Speech and Signal Processing (ICASSP)}, pages 2747--2751. IEEE, 2019.

\bibitem{kaselimi2020context}
Maria Kaselimi, Nikolaos Doulamis, Athanasios Voulodimos, Eftychios
  Protopapadakis, and Anastasios Doulamis.
\newblock Context aware energy disaggregation using adaptive bidirectional lstm
  models.
\newblock {\em IEEE Transactions on Smart Grid}, 2020.

\bibitem{mauch2015new}
Lukas Mauch and Bin Yang.
\newblock A new approach for supervised power disaggregation by using a deep
  recurrent lstm network.
\newblock In {\em 2015 IEEE Global Conference on Signal and Information
  Processing (GlobalSIP)}, pages 63--67. IEEE, 2015.

\bibitem{wang2019nonintrusive}
Ke~Wang, Haiwang Zhong, Nanpeng Yu, and Q~Xia.
\newblock Nonintrusive load monitoring based on sequence-to-sequence model with
  attention mechanism.
\newblock In {\em Zhongguo Dianji Gongcheng Xuebao/Proceedings of the Chinese
  Society of Electrical Engineering}, volume~39, pages 75--83, 2019.

\end{thebibliography}

\end{document}